\documentclass{emulateapj}
\usepackage{mathptmx,courier}

\newcommand{\code}[1]{\textsc{#1}}
\newcommand{\Flash}{\code{Flash}}

\newcommand{\Umax}{\ensuremath{U_{\mathrm{max}}}}
\newcommand{\Mach}{\ensuremath{\mathsf{Ma}}}

\newcommand{\MCO}{\ensuremath{M_{\mathrm{CO}}}}

\newcommand{\rhowd}{\ensuremath{\rho_{\mathrm{WD}}}}

\newcommand{\cs}{\ensuremath{c_{\mathrm{s}}}}

\newcommand{\nuclei}[2]{\ensuremath{^{#1}\mathrm{#2}}}
\newcommand{\C}{\nuclei{12}{C}}
\newcommand{\Ox}{\nuclei{16}{O}}

\newcommand{\unitspace}{\ensuremath{\,}}
\newcommand{\usp}{\unitspace}
\newcommand{\unitstyle}[1]{\ensuremath{\mathrm{#1}}}
\newcommand{\power}[2]{\ensuremath{{#1}^{#2}}}
\newcommand{\unit}[2]{\ensuremath{#1\unitspace\unitstyle{#2}}}

\newcommand{\centi}{\unitstyle{c}}
\newcommand{\kilo}{\unitstyle{k}}

\newcommand{\cm}{\centi\meter}
\newcommand{\gram}{\unitstyle{g}}

\newcommand{\meter}{\unitstyle{m}}
\newcommand{\second}{\unitstyle{s}}

\newcommand{\Msun}{\ensuremath{M_\odot}}
\newcommand{\yr}{\unitstyle{yr}}        

\newcommand{\grampercc}{\gram\unitspace\power{\cm}{-3}} 
\newcommand{\ergs}{\unitstyle{ergs}}
\newcommand{\ergspersecond}{\ergs\unitspace\power{\second}{-1}}
\newcommand{\ergspergram}{\ergs\unitspace\power{\gram}{-1}}
\newcommand{\ergspergrampersecond}{\ergspergram\unitspace\power{\second}{-1}}
\newcommand{\km}{\kilo\meter}
\newcommand{\Msunperyr}{\Msun\unitspace\power{\yr}{-1}}

\slugcomment{to appear in \textsc{the Astrophysical Journal}}

\shorttitle{Enrichment in Classical Novae}
\shortauthors{Alexakis et al.}

\begin{document}

\title{On Heavy Element Enrichment in Classical Novae}

\author{
A.~Alexakis\altaffilmark{1,2},
A~C.~Calder\altaffilmark{1,3},
A.~Heger\altaffilmark{1,4,5},
E.~F.~Brown\altaffilmark{1,3},
L.~J.~Dursi\altaffilmark{1,3},
J.~W.~Truran\altaffilmark{1,3,4},
R.~Rosner\altaffilmark{1,2,3,4},
D.~Q.~Lamb\altaffilmark{1,3,4},
F.~X.~Timmes\altaffilmark{1,3},
B.~Fryxell\altaffilmark{1,4},
M.~Zingale\altaffilmark{6},
P.~M.~Ricker\altaffilmark{7,8},
and
K.~Olson\altaffilmark{1,9}
}

\email{alexakis@flash.uchicago.edu}

\altaffiltext{1}{Center for Astrophysical Thermonuclear Flashes,
                 The University of Chicago,
                 Chicago, IL  60637}
\altaffiltext{2}{Department of Physics,
                 The University of Chicago,
                 Chicago, IL  60637}
\altaffiltext{3}{Department of Astronomy \& Astrophysics,
                 The University of Chicago,
                 Chicago, IL  60637}
\altaffiltext{4}{Enrico Fermi Institute,
                 The University of Chicago,
                 Chicago, IL  60637}
\altaffiltext{5}{Los Alamos National Laboratory, 
                 Theoretical Astrophysics Group,
                 Los Alamos, NM 87545}
\altaffiltext{6}{Department of Astronomy and Astrophysics, 
                 The University of California, Santa Cruz,
                 Santa Cruz, CA 95064}
\altaffiltext{7}{Department of Astronomy,
                 University of Illinois, 
                 Urbana, IL 61801}
\altaffiltext{8}{National Center for Supercomputing Applications,
                 Urbana, IL 61801}
\altaffiltext{9}{UMBC/GEST Center,
                 NASA/GSFC,
                 Greenbelt, MD 20771}

\begin{abstract}
  Many classical nova ejecta are enriched in CNO and Ne.  Rosner et al.\ 
  recently suggested that the enrichment might originate in the resonant
  interaction between large-scale shear flows in the accreted H/He
  envelope and gravity waves at the interface between the envelope and
  the underlying C/O white dwarf.  The shear flow amplifies the waves,
  which eventually form cusps and break.  This wave breaking injects a
  spray of C/O into the superincumbent H/He.  Using two-dimensional
  simulations, we formulate a quantitative expression for the amount of
  C/O per unit area that can be entrained, at saturation, into the H/He.
  The fraction of the envelope that is enriched depends on the
  horizontal distribution of shear velocity and the density contrast
  between the C/O white dwarf and the H/He layer but is roughly
  independent of the vertical shape of the shear profile.  Using this
  parameterization for the mixed mass, we then perform several
  one-dimensional Lagrangian calculations of an accreting white dwarf
  envelope and consider two scenarios: that the wave breaking and mixing
  is driven by the convective flows; and that the mixing occurs prior to
  the onset of convection.  In the absence of enrichment prior to
  ignition, the base of the convective zone, as calculated from
  mixing-length theory with the Ledoux instability criterion, does not
  reach the C/O interface.  As a result, there is no additional mixing,
  and the runaway is slow.  In contrast, the formation of a mixed layer
  during the accretion of H/He, prior to ignition, causes a more violent
  runaway.  The envelope can be enriched by $\lesssim 25\%$ of C/O by
  mass (consistent with that observed in some ejecta) for shear
  velocities, over the surface, with Mach numbers $\lesssim 0.4$.
\end{abstract}


\section{Introduction}

Classical novae are a manifestation of thermonuclear runaways in
accreted hydrogen shells on the surfaces of white dwarf stars in close
binary systems \citep[see, e.g., the review by][]{gehrz98}.  Compelling
observational data indicate that the material ejected by some classical
novae can be significantly enriched in C, N, O, and Ne, by $\gtrsim
30\%$ by mass (\citealt{livio94,gehrz98}).  Since the abundance of CNO
catalysts constrains the rate of energy release, such high levels of CNO
enrichment are required in the fastest novae \citep{truran82}, for which
the hydrogen burning reactions (via the CNO cycle) increase the entropy
of the envelope faster than it can adjust.  It was recognized early that
if some of the underlying white dwarf matter\footnote{By ``white
  dwarf,'' we mean the predominantly C/O substrate; we shall refer to
  the accumulated H/He layer as the ``atmosphere'' or ``envelope.''}
could be mixed into the accreted envelope prior to the final stages of
the thermonuclear runaway, then the explosion would be more energetic
and the ejecta more enriched in CNONe elements
\citep{starr72}. One-dimensional models that best reproduce observations
typically accrete material ``seeded'' with a super-solar composition
\citep[e.g.,][]{hernanz96,starr98}.

The question of how the accreted envelope is enriched has challenged
theory for several decades \citep[see, e.g. the review by][]{livio94}.
For very slow accretion ($\lesssim 10^{-10}\usp\Msunperyr$), the
downward diffusion of H into the underlying C/O or O/Ne layers
\citep{prialnik84,prialnik85} could trigger ignition in the H diffusive
tail and drive convective mixing of heavy elements into the envelope
during the early stages of runaway.  It is unclear, however, whether
there is time at higher accretion rates for this process to be relevant.
The possibility of mixing via convective overshoot was first considered
by \citet{woosley86}.  Numerical simulations of convective penetration
and mixing at the core-envelope interface have been carried out in both
two dimensions \citep*{glasner97, kercek98} and three \citep*{kercek99}.
From comparisons of two- and three-dimensional simulations and a careful
resolution study, \citeauthor{kercek98}~(\citeyear{kercek98},
\citeyear{kercek99}) concluded that convective penetration would not
significantly enrich the accreted envelope.

The general problem of shear mixing was considered by many
\citep{kippenhahn78,macdonald83,fujimoto88}, but to date no
self-consistent calculations of these effects have been performed.
\citet{kippenhahn78} considered shearing from differential rotation in
the envelope and estimated the amount of mass mixed by assuming that the
envelope was marginally stable according to the Richardson criterion.
\citet{macdonald83} considered non-axisymmetric perturbations and the
resulting redistribution of matter and angular momentum.
\citet{fujimoto88} emphasized the role of barotropic and baroclinic
instabilities in transporting angular momentum through the envelope.  A
variation on this was envisaged by \citet{sparks87} and
\citet{kutter89}, who suggested that convection just prior to the
runaway would transport angular momentum inward and lead to a large
horizontal shear above the C/O white dwarf.

While these studies examined the generation of shear in the envelope,
the details of how C/O is mixed into the envelope and the quantity of
mass mixed, as functions of the shear profile, were not resolved.  In
this paper, we examine a promising mechanism for effecting the
entrainment: a resonant interaction between large-scale shear flows in
the accreted envelope and interfacial gravity waves \citep{rosner01}.
The greater compositional buoyancy in the C/O white dwarf means that the
interface sustains gravity waves.  \citet{Miles57} showed that in the
presence of a shear flow (i.e., a ``wind''), gravity waves with a group
velocity matching a velocity in the shear flow are resonantly amplified.
These waves eventually form cusps and break.  When the waves break, they
inject, analogously to ocean waves, a spray of C/O into the H/He
atmosphere.

The source of the shear could arise from a number of mechanisms, such as
those described above.  A self-consistent calculation of the accretion
boundary layer and the wind profile in the atmosphere is far beyond the
scope of current simulations.  Indeed, as we explain in
\S~\ref{sec:gravitational_waves}, the resolution needed to simulate the
interaction is so fine as to prohibit the direct simulation of this
effect within a macroscopically large (i.e., the entire atmosphere of
the white dwarf) computational domain.  What this paper provides is a
scaling relation (eq.~[\ref{eq:MCO}]) for the mixing rate per unit area,
for a given wind velocity.  This relation may be used to construct a
sub-grid model for large-scale multi-dimensional simulations, which can
in turn address how matter and angular momentum are transported over the
surface.

In \S~\ref{sec:gravitational_waves} we present the two-dimensional
models and demonstrate how the mixed mass depends on the velocity of the
flow.  We then, in \S~\ref{sec:nova1d}, incorporate this relation into a
one-dimensional simulation and discuss two scenarios: that mixing only
occurs if the convective zone reaches the C/O interface; and that the
mixing occurs prior to the onset of convection.  We implement these
results as a sub-grid model in one-dimensional novae simulations
performed with a stellar evolution code and discuss the implications, in
\S~\ref{sec:implications}, for the ejected mass of C/O.  Because the
shear in the envelope prior to the onset of convection is unknown, we
compute the accretion and runaway for several different masses of the
mixed layers.

\section{Breaking Gravity Waves}
\label{sec:gravitational_waves} 

To calculate the amount of mass mixed by wave breaking as a function of
time, we performed a suite of numerical simulations of wind-driven
gravity waves.  The initial configuration consists of two stably
stratified layers of compressible fluids in hydrostatic equilibrium in a
uniform gravitational field $g$.  The upper fluid is composed of H/He
and the lower fluid is composed of C/O.  The equation of state is an
ideal gamma-law of internal energy $\rho\epsilon = P/(\gamma-1)$.  We
choose $\gamma = 5/3$ and for the initial condition we construct an
isentropic envelope in hydrostatic equilibrium. The density is then
given by $\rho(z)=\rho_i[1-(\gamma-1)z/H]^{1/(\gamma-1)}$, where
$\rho_{i=1}$ is the density of the upper fluid just above the interface,
$\rho_{i=2}$ is the density of the lower fluid just below the interface,
$H=\gamma^{-1}\cs^2/g$ is the pressure scale height, and \cs\ is the
adiabatic sound speed. This choice makes the Brunt-V\"ais\"al\"a
frequency zero, i.e., there are no internal gravity waves within each of
the two layers, which isolates the interface modes for study.

We impose a wind in the upper fluid with velocity
\begin{equation}\label{eq:profile}
  U(z)= \Umax (1-e^{-z/\delta}),
\end{equation}
where $z$ is the vertical direction and $z=0$ corresponds to the
location of the initial interface; \Umax\ is the asymptotic maximum
velocity at $z\gg \delta$, and $\delta$ is the length scale of the shear
boundary layer.  The form of the wind profile (eq.~[\ref{eq:profile}])
is chosen to model the boundary layer formed at the surface of the white
dwarf, and is motivated by studies of the wind-wave resonant instability
in oceanography \citep{Miles57}.  This instability is the principal
mechanism that drives gravity waves and air-water mixing in oceans. In
the limit $\delta \to 0$, one recovers the Kelvin-Helmholtz wind
profile.  Without a study of angular momentum transport in the accreting
envelope, we do not know the form of the wind profile and in this work
investigate the mixing for a range of $\delta$, with $\delta/H \ll 1$.

\citet*{alexakis02} performed a linear stability analysis of the wave
amplification for the novae problem, and in a forthcoming paper 
(Alexakis et al., in preparation) we will present extensive numerical
studies of wave-wind interactions.  Here we apply those results to the
conditions at the base of the white dwarf envelope.  The simulations are
performed using the \Flash\ code \citep{fryxell00,calder00,calder02}, a
parallel, multi-dimensional, adaptive-mesh code for simulating
compressible, inviscid flow.  The computational grid was a uniform mesh
of resolution $1024\times1024$ with periodic boundary conditions along
the sides of the box and hydrostatic boundaries \citep{zingale02} at the
top and bottom.  The interface is perturbed with specified modes and a
velocity perturbation added to the surface of the WD representing the
motion of the gravity waves.  The wind profile is specified as an
initial condition and in particular is not forced.  For this study, we
set the Mach number $\Mach = \Umax/\cs = 0.5$ and consider four values
of $\delta/H$: 0.005, 0.01, 0.02, and 0.04.  These values vary the
reciprocal Froude number $(\delta g/\Umax^2)^{1/2} =
(\delta/H)/(\gamma\Mach^2)$, which measures the available kinetic energy
for mixing.

\begin{figure}
\includegraphics[width=\columnwidth]{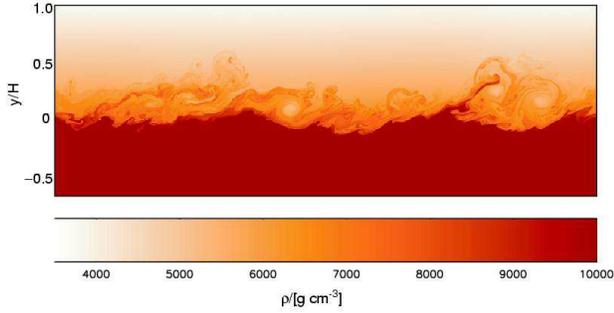}
\caption{Breaking CO waves, as determined by simulations in two
  dimensions.  Gravity points towards the bottom of the figure, with the
  vertical distance $y$ in units of the pressure scale height $H$, as
  evaluated just above the interface.  The color scale indicates the
  mass density in units of \grampercc.
\label{fig:break_layer}}
\end{figure}

\begin{figure}
\includegraphics[width=\columnwidth]{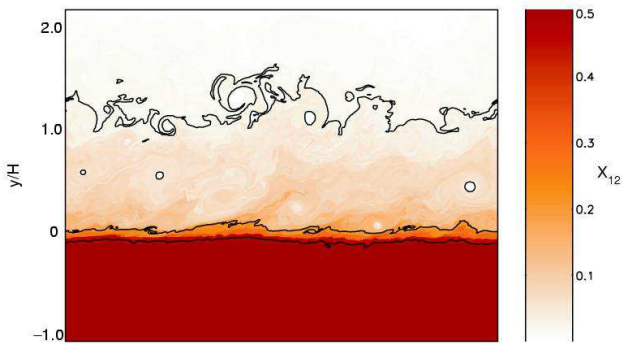}
\caption{Mass fraction of \C\ for $\delta/H = 0.04 $ after $t =
    3500 \delta/U $.  The vertical dimension is scaled to the pressure
    scale height $H$ as evaluated just above the interface.  The
    contours for \C\ mass fractions of, from the top, 0.02, 0.20, and
    0.49.
\label{fig:c12_contours}}
\end{figure}

The unstable modes have wavenumbers \citep{alexakis02} $k \gtrsim
g/\Umax^2 (1-\rho_1/\rho_2)$. We chose the size of our computational
domain to be roughly equal to this maximum wavelength, so that we have
1024 grid points along the largest possible wave. In practice, smaller
waves become dominant, with the wavelength of the dominant mode
depending on the wind profile.  In all cases this dominant mode is
well-resolved.  Figure~\ref{fig:break_layer} shows fully developed waves
breaking, generation of the ``spray", and mixing of the white dwarf
substrate up into the atmosphere.  Mixing is not only effected by the
overturning of the wave but also by the wind breaking off the tip of the
wave.  Figure~\ref{fig:c12_contours} shows contours (\emph{black lines})
of the carbon mass fraction at 0.49, 0.20, and 0.02, respectively, for
the simulation depicted in Figure~\ref{fig:break_layer}, but at a later
time. The contour at 0.49 corresponds to the carbon mass fraction of the
underlying white dwarf, while the smaller contour values indicate how
far outward into the accreted material the white dwarf substrate is
mixed. The contour at 0.2 delimits the region that contains most of the
initial enrichment.

\begin{figure}
\includegraphics[width=\columnwidth]{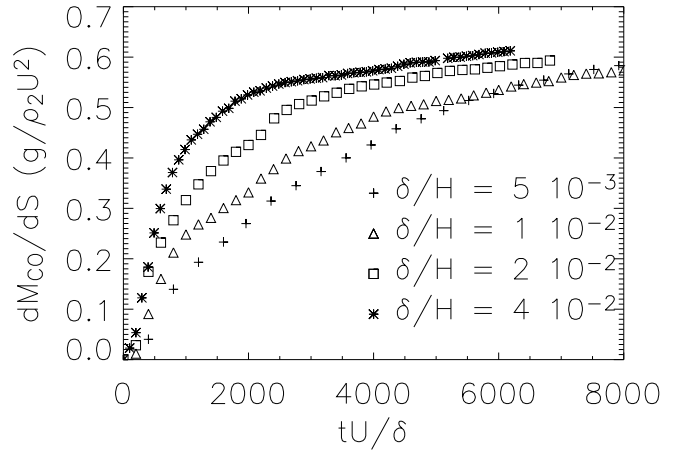}
\caption{The mixed mass of C/O, per unit area, as a function of time.
  This was computed by averaging over the horizontal direction in the
  simulations.  Time is scaled to $\delta/\Umax$ and \MCO\ is scaled to
  $\rho_2\Umax^2/g$ (see eq.~[\ref{eq:MCO}]).
  Four different values of $g\delta/\Umax^2$ were used, 0.01, 0.02,
  0.04, and 0.09.
\label{fig:mixc12}
}
\end{figure}

Figure~\ref{fig:mixc12} shows the surface mass density (the mass of C/O
in the mixed layer per unit area), $d\MCO/dS$, averaged over the
horizontal direction. The mixed layer is defined here as the region in
which the carbon mass fraction is between 0.49 and 0.01.  The amount of
mixed mass depends on these delimiters; reducing the upper limit to 0.4
decreases the mixed mass by less than a factor of 2.  The C/O is mixed
rapidly until it saturates; further mixing occurs on diffusive
timescales.  Our two-dimensional simulations show, for the range of
parameters examined, that the total mass of white dwarf material that
becomes mixed is independent of the lengthscale $\delta$.  The rate of
mixing, i.e., the initial slopes of the curves in
Figure~\ref{fig:mixc12}, does show, however, some dependence on
$\delta$.  Using dimensional analysis and the numerical results, we find
that for a fixed density ratio $\rho_1/\rho_2 = 0.6$ (appropriate for
the degenerate plasma prior to ignition) the total mass per unit area,
$d\MCO/dS$, mixed into the accreted H/He saturates at
\begin{equation}\label{eq:MCO}
  \frac{d\MCO}{dS} = \alpha \frac{\Umax^2}{g}\rhowd ,
\end{equation}
where $\alpha$ is a non-dimensional constant\footnote{In general,
  $\alpha$ depends on $\rho_2/\rho_1$ and possibly \Mach.  A parameter
  study, outside the scope of this paper, is necessary to determine this
  dependence.}  that we determine from the simulations to be
$\alpha\simeq 0.6$ (see Figure~\ref{fig:mixc12}).  The timescale to reach
saturation is far shorter than the timescale of either the accretion
phase ($> 10^4\usp\yr$) or the pre-peak convective phase ($\sim
100\usp\yr$) of a typical classical nova.  For example, if $\delta/H =
0.01$ and $\Mach = 0.5$, then the saturation timescale is of order
seconds for our nova setup.

\section{One-Dimensional Nova Models}
\label{sec:nova1d} 

We now incorporate our simulations of the wave breaking and mixing into
simulations of the thermonuclear burning of a nova.  In order to explore
the global properties of this local mixing mechanism, we compute several
one-dimensional models of novae with a modified version of the
\code{KEPLER} stellar evolution code \citep*{weaver78} that includes a
large network of $\sim100$ light isotopes \citep{rau02} implicitly
coupled into the stellar structure solver.  This allows us to follow
throughout the entire convective envelope the radioactive decays of
nuclei that are formed at the base of the burning zone.  The underlying
white dwarf has a mass of $1.0\usp\Msun$, a radius of $5000\usp\km$, a
luminosity of $10^{31}\usp\ergspersecond$, and is composed of a
50\%/50\% \C/\Ox\ mixture.  We retain the outer $0.005\usp\Msun$ on the
computational grid to follow the thermal inertia of the outer WD layers
and allow the model to relax until it has a constant luminosity before
we start the accretion.  The accretion rate is $10^{-9}\usp\Msunperyr$,
and in all but one model (see item 3 below), the accreted material
consists of 70\,\% \nuclei{1}{H}, 28.7\,\% \nuclei{4}{He}, 0.3\,\% \C,
0.1\,\% \nuclei{14}{N}, and 0.9\,\% \Ox, by mass.  Convection is modeled
using the Ledoux criterion for stability and mixing length theory.  We
assume no convective overshooting and that semiconvection is too slow to
cause mixing.

We investigate three scenarios, as summarized in Table~\ref{1dtab}, for
generating the gravity wave induced mixed layer.

\begin{table}
  \centering
  \caption{Properties of the 1D nova models\label{1dtab}}
  \begin{tabular}{llllllll}
    \hline\hline
    \noalign{\smallskip}
    Case
    & $M_{\mathrm{enrich}}$
    & $M_{\mathrm{envel}}$
    & $\dot{E}_{\mathrm{nuc},\max}$
    \\
    & ($M_{\odot}$)
    & ($M_{\odot}$)
    & (\ergspersecond)
    \\
    \noalign{\smallskip}
    \hline
    \noalign{\smallskip}
    no pre-enrichment      &0                  &$6.93\times10^{-5}$&$6.68\times10^{42}$\\ 
    \hline
    wave pre-enrichment&$1.26\times10^{-9}$&$7.08\times10^{-5}$&$7.24\times10^{42}$\\ 
    wave pre-enrichment&$1.26\times10^{-8}$&$7.05\times10^{-5}$&$7.65\times10^{42}$\\ 
    wave pre-enrichment&$1.26\times10^{-7}$&$6.78\times10^{-5}$&$7.71\times10^{42}$\\ 
    wave pre-enrichment&$1.31\times10^{-6}$&$5.42\times10^{-5}$&$1.27\times10^{43}$\\ 
    wave pre-enrichment&$3.77\times10^{-6}$&$4.67\times10^{-5}$&$2.68\times10^{43}$\\ 
    wave pre-enrichment&$1.16\times10^{-5}$&$4.85\times10^{-5}$&$1.26\times10^{44}$\\ 
    \hline
    enriched accretion &(50\,\%)           &$2.08\times10^{-6}$&$5.28\times10^{40}$\\ 
    \noalign{\smallskip}
    \hline\hline
  \end{tabular}
\end{table}

\begin{enumerate}
\item In the first scenario, the only shearing considered is that from
  the convective cells driving a wind at the interface between the
  H-rich atmosphere and the C/O substrate.  This is the scenario
  envisaged by \citet{rosner01}, in which the C/O, after being
  entrained, is then distributed throughout the convective zone.
  Figure~\ref{fig:plain} summarizes the evolution of the accreted layer.
  After accretion for about $6.9\times10^4\usp\yr$ a convective zone
  forms about $2.5\times10^{-5}\usp\Msun$ above the WD interface.  The
  total envelope mass accreted at ignition is $\approx 7\times
  10^{-5}\usp\Msun$, in good agreement with the estimates of
  \citet{fujimoto82a,fujimoto82b}.  Within $\sim100\usp\yr$ from the
  onset of convection, the convective zone extends upward to the surface
  and downward to about $5\times10^{-6}\usp\Msun$ above the interface,
  at which time the runaway reaches its peak rate of energy generation.
  This peak phase evolves on a timescale of $\sim\unit{1}{h}$, but
  the convection does not reach the WD interface.  Only days later, when
  the nova envelope is already significantly expanding, does the burning
  layer reach the WD.  This downward movement, in mass, of the burning
  layer is mediated by heat conduction into these deeper H-rich layers.
  The convective zone also moves deeper, but never quite reaches the WD
  interface\footnote{At the very bottom of the H shell the nuclear
    luminosity from H burning can never become large enough to drive
    convection, as it goes to zero at the interface; the only
    possibility for mixing to occur is if a steep temperature gradient
    arises from the thermal inertia of the heated WD core below an
    expanding H envelope.}.  As a result, there is no injection of C/O
  into the H-rich envelope, by construction.

  \begin{figure}
  \includegraphics[angle=90,width=\columnwidth]{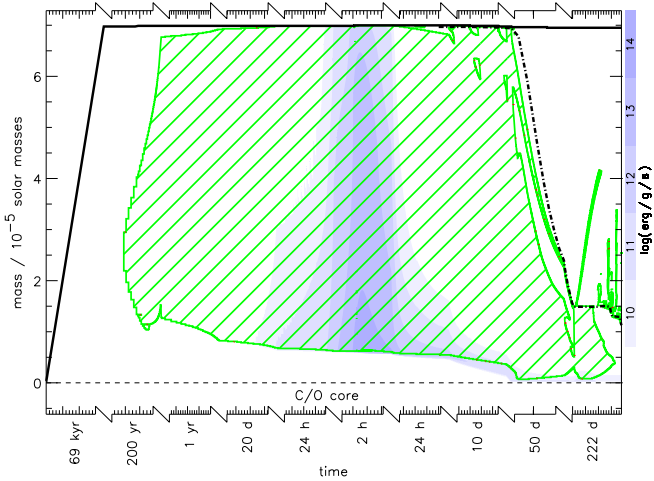}
  \caption{ Kippenhahn diagram of a nova without enrichment.  The
    $x$-axis indicates time intervals for the different evolution stages,
    and the $y$-axis gives the mass above the C/O WD substrate.
    \emph{Green hatching} (framed by a \emph{green line}) indicates
    convection, \emph{blue shading} indicates nuclear energy generation
    for which each level of darker blue denotes an increase by one order
    of magnitude, starting at $10^{10}\usp\ergspergrampersecond$.  The
    \emph{thick black line} shows the total mass of star (including
    ejecta), increasing because of accretion; the \emph{dash-dotted line}
    indicates the mass outside of $10^{12}\usp\cm$; and the \emph{dashed
      line} marks the interface between the white dwarf C/O substrate and
    the accreted layers.
    \label{fig:plain}
  }
  \end{figure}

    The envelope expands to large radii ($>10^{12}\usp\cm$, large enough
    to engulf the secondary) by the end of the simulation, but the
    result will only be a slow nova, in rough agreement with the
    semi-analytical calculations of \citet{fujimoto82a,fujimoto82b}.  We
    do not find any contact of the convective region with the WD
    substrate; as a result, we do not expect that any convection-driven
    wave mixing would occur, unless there is a large redistribution of
    angular momentum in the accreted envelope \citep{sparks87,kutter89}.
    
  \item In the second scenario, the shearing originates from a wind at
    the base of the accreted envelope blowing across the underlying C/O
    substrate.  We assume that the wind persists throughout the H/He
    layer with velocity sufficient to drive mixing on a timescale much
    less than that to accrete a critical mass of fuel.  In this case the
    mixed layer is generated prior to ignition.  We also assume a linear
    (in Lagrangian mass coordinate) gradient in the mass fraction of C/O
    between the WD and the accreted envelope.  Because the shear
    profile, and hence the amount of mass mixed, is unknown, we consider
    a range of mixed masses (see Table~\ref{1dtab}).  Our ignorance of
    the fluid motions in the envelope and substrate prevents us from
    saying where and how the accreted material spreads over the surface,
    and so we cannot determine the actual shear at the base of the
    envelope.  In all cases that we consider, the shear velocity is
    subsonic, which is the range of validity of the 2-d simulations.
    For such velocities, the mixed layer is always thinner than a
    pressure scale height (eq.~[\ref{eq:MCO}]).
    
    Figure~\ref{fig:rich} shows the case with the largest
    pre-enrichment: the mixed material comprises
    $4.6\times10^{28}\usp\gram$ (i.e., a total of
    $2.3\times10^{28}\usp\gram$ of WD material is being mixed with
    H-rich material).  This is about 25\% of the envelope mass, which is
    less than that generated by the calculation described in
    section~\ref{sec:gravitational_waves}.  From
    equation~(\ref{eq:MCO}), a 25\% enrichment corresponds to $\Mach =
    0.4$ if the velocity were uniform over the surface.  We note that
    this velocity is much less than Keplerian, $\Umax \approx 0.05
    (GM/R)^{1/2}$, as expected in the envelope well below the accretion
    disk boundary layer \citep[see, e.g.,][]{popham95}.  If the velocity
    were not uniform over the surface, then a higher maximum velocity
    would be required to inject the same percentage of the total
    envelope mass.

    \begin{figure}
      \includegraphics[angle=90,width=\columnwidth]{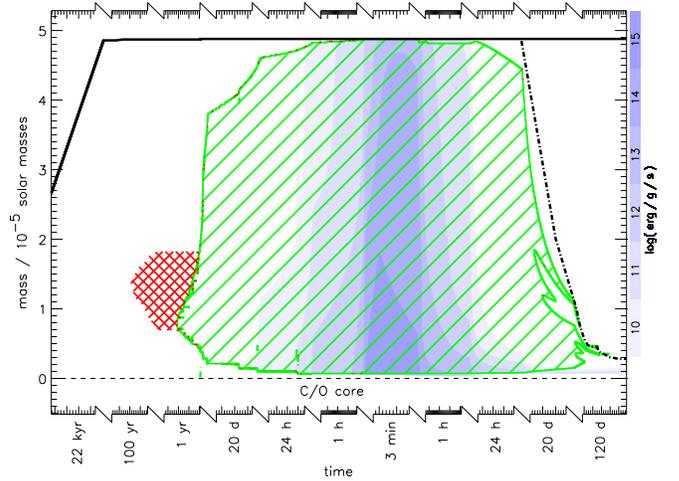}
      \caption{ Same as Fig.~\ref{fig:plain} but for a model in which the
        inner $M_{\mathrm{mix}}=4.6\times10^{28}\usp\gram$ are enriched in C/O
        with a linear composition gradient (with respect to the Lagrangian
        mass coordinate) between the WD composition (C/O) at the base and the
        accretion composition (solar) at the upper edge. Note that the
        convective zone does not reach the interface with the WD substrate,
        and that a significant semi-convective region, indicated by
        \emph{red hatching}, develops prior to the onset of convection.
        \label{fig:rich}
      }
    \end{figure}
    
    The mixed layers are added to the surface using the same accretion
    rate as the rest of the envelope ($10^{-9}\usp\Msunperyr$); to
    prevent a runaway while these layers are being added, we suspend
    nuclear energy generation during the accretion of the first
    $5\times10^{28}\usp\gram$.  About 50\usp\yr\ prior to peak energy
    generation a semiconvective region forms in the outer 60\,\% (by
    mass) of the enriched layer; about 8 months before the peak a
    convection zone starts at the base of this semiconvective layer (at
    $7\times10^{-6}\usp\Msun$ above the WD interface) that extends
    upward to the surface and downward to about
    $5\times10^{-7}\usp\Msun$ above the interface.  In contrast to the
    first case, the pre-mixed envelope with the largest enriched mass
    ignites at a smaller accumulated mass, $\approx 5\times
    10^{-5}\usp\Msun$.

    The energy generation in the enriched layer is dominated by
    $\C(\mathrm{p},\gamma)\nuclei{13}{N}(\beta^+)\nuclei{13}{C}$.  The
    peak energy generation timescale is about \unit{3}{min}---20 times
    faster than in the first case---with a specific energy generation
    rate about 20 times higher.  Figure~\ref{fig:enuc} shows the
    specific nuclear energy generation in the accreted envelope, with
    the time centered about the peak energy generation rate.  The larger
    rate of energy generation roughly corresponds to the $\sim 20$ times
    greater metal enrichment.  In this scenario the entire 25\,\% of
    pre-enriched C/O material is spread throughout the hydrogen
    envelope during the runaway.  This leads to a a corresponding
    enrichment in the ejecta.  The ejecta are unbound, i.e., they have a
    positive velocity at very large radii.

    \begin{figure}
      \includegraphics[angle=0,width=\columnwidth]{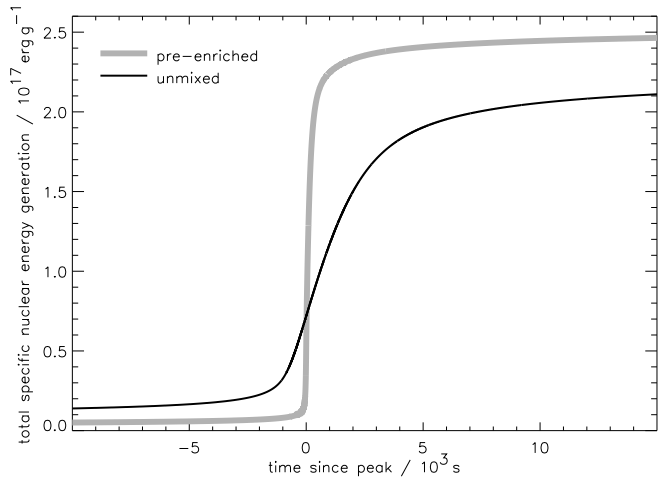}
      \caption{ Time-integtated average specific nuclear energy generation
        of the accreted enevelope.  The curves are shifted so that the zero in
        time corresponds to the peak energy generation rate.  The gray curve
        indicates the model with the highest wave-induced pre-enrichment (see
        Tab.~\protect\ref{1dtab} and Fig.~\protect\ref{fig:rich}); the black
        curve indicates the model without prior enrichemnt
        (Fig.~\protect\ref{fig:plain}).  The slope of a curve is proportional
        to the average specific energy generation rate in the envelope.
        \label{fig:enuc}
      }
    \end{figure}

    As in the first case, the convection never reaches the WD interface
    (although the base of the convective zone is much deeper), so there
    is no additional convection-induced wave mixing of the C/O substrate
    with the hydrogen-rich envelope.  We did experiment with
    pre-runaway, pre-enriched layers as small as 1/10,000 the mass used
    in the case described above (Table~\ref{1dtab}).  As before, we
    disregarded nuclear burning for the first $10^{28}\usp\gram$ of
    accreted material, with the exception of the case with
    $3.77\times10^{-6}\usp\Msun$ of enrichment, where we took twice that
    mass to accommodate the entire enriched layer.  The difference in
    masses for which nuclear burning is suspended is the reason for the
    case with the largest pre-enriched mass,
    $1.15\times10^{-5}\usp\Msun$ having a slightly larger envelope mass
    at runaway than that of the case with $3.77\times10^{-6}\usp\Msun$
    of enriched mass.

    In none of these cases did the convection reach the WD interface.
    Eventually, for the case of an enriched mass of
    $\lesssim10^{-9}\usp\Msun$, the runaway behaves as in the
    non-enriched case: the energy release rate in the mixed layers is
    too small to drive convection above and heat conduction cools it
    efficiently so that the runaway does not occur in its vicinity.  The
    slightly higher envelope masses obtained in the limit of small
    enrichment is likely to be an artifact of our suspension of nuclear
    energy generation for the first fraction of the accretion phase.

  \item In a third scenario we assume accretion of material enriched to
    a mass fraction of 50\,\% in C/O.  Allowing for nuclear energy
    generation instantaneously in this material leads to a rapid runaway
    after only a thin layer has been accumulated at the surface of the
    star.  If this magnitude of enrichment was due to instantaneous
    mixing with a colder WD substrate, then a lower temperature in the
    enriched material and thus a significantly later runaway could
    result.  We include this run to demonstrate how qualitatively
    different the accretion of pre-enriched material behaves as compared
    to first accreting unenriched material and then, only later after it
    has settled, mixing the accreted layers with the C/O substrate.

\end{enumerate}

\section{Implications and Summary}
\label{sec:implications} 

Using a constant wind profile that blows across the surface of a white
dwarf, we have performed a two-dimensional parameter study of the
mechanism proposed by \cite{rosner01}.  Our primary results of the
mixing rate and the maximum mixed mass, equation~(\ref{eq:MCO}), suggest
that this process can mix about $10^{-6}$--$10^{-5}\usp\Msun$ of the
underlying C/O into the hydrogen-rich envelope.  From this, we
investigated two scenarios: that the wind is the result of convection,
and that the wind originates during the accretion phase.

We find that if no enrichment occurs prior to the onset of convection,
then the convective zone does not reach the C/O interface, and no
additional mixing occurs (in the one-dimensional model) in the absence
of convective overshoot.  The result in this case will be a slow nova,
with little enrichment of the ejecta.  In contrast, an envelope with a
mixed layer at the C/O interface, consistent with the scalings from
high-resolution numerical studies, provides a more violent runaway and
the ejecta are enriched by $\sim 25\%$ in C/O, consistent the enrichment
observed in some nova ejecta.  Such an event does require a strong shear
velocity, but the saturation amount of mass mixed is roughly independent
of the shape of the shear profile, so long as its thickness is much less
than a pressure scale height.  Our runs, outlined in Table~\ref{1dtab},
show that when the metals are concentrated at the base of the accreted
envelope, the amount of mass accreted prior to runaway is \emph{larger}
than if the C/O were uniformly distributed over the envelope, i.e., if
the white dwarf were to accrete material with a supersolar metallicity.
The reason is that the opacity of the envelope is larger than when the
C/O are concentrated at the base.

There are a number of issues that we have not yet addressed or
investigated. First, our results might depend on the dimensionality of
the system. We would therefore expect that although the generation of
the gravity waves is captured by two-dimensional dynamics, the energy
cascade of the waves and the advection of spray and vorticity will be
different in a three dimensions.  Second, this work investigates only
the density ratio for an accreted envelope in thermal equilibrium.
During the early phase of accretion and during the runaway the density
ratio will likely be different.  Third, although our parameter study
covers more than one order of magnitude in $\delta$, it is important to
know if our results still hold at even larger values.  As $\delta$
becomes a respectable fraction of $H$, we expect that the stratification
of the envelope, and in particular the effect of a non-zero
Brunt-V\"ais\"al\"a frequency, will become important.  All of these
differences can affect the mass and thickness of the mixed layer, and
the investigation of these issues is the subject of ongoing work.

In our one-dimensional nova simulations, we assumed that a convective
roll will only entrain the mixed layer if the base of the convective
zone (as computed from a mixing-length formalism) reaches the interface.
Future multidimensional studies can inform us as to how convection
interacts with the interfacial gravity waves, and to whether this
assumption is in fact correct.  Moreover, our sub-grid model, when
incorporated into a one-dimensional calculation, does not account for
spatial variations of the wind.  Such a variation would occur, for
example, if the fuel accretes non-uniformly over the surface of the star
and then spreads.  In reality the amount of mass mixed will vary over
the surface of the star.  Our one-dimensional calculations must of
necessity take the mixed mass as a free parameter.  Because the amount
of C/O entrained depends quadratically on the velocity in the wind
profile, our calculation underscores the need for detailed simulations
of the shear profile of an accreting white dwarf.  These issues are
clearly important and are the next steps to pursue.

\acknowledgements

We thank Lars Bildsten and Stan Woosley for helpful discussions.  This
work is supported in part by the U.S. Department of Energy under Grant
No.~B341495 to the Center for Astrophysical Thermonuclear Flashes at the
University of Chicago.  A. H. is supported by the Department of Energy
under contract W-7405-ENG-36.  L. J. D. acknowledges support by the
Krell Institute CSGF.  M. Z. is supported by DOE grant number
DE-FC02-01ER41176 to the Supernova Science Center/UCSC.  P.~M.~R.\ 
acknowledges support from the University of Illinois at Urbana-Champaign
and the National Center for Supercomputing Applications (NCSA).  K.
Olson acknowledges partial support from NASA grant NAS5-28524.

\end{document}